\documentclass[aps,prl,reprint,
superscriptaddress,
floatfix,
nofootinbib,
longbibliography]{revtex4-2}
\usepackage{upgreek}
\usepackage{graphicx}
\usepackage{amsmath}
\usepackage{dcolumn}
\usepackage{bm}
\usepackage{hyperref}
\usepackage{svg}
\usepackage{color}
\usepackage{braket}

\usepackage{soul}

\begin{document}

\title{Near-deterministic loading of optical tweezer arrays via repulsive barricade potentials}

\author{Archie C. Baldock}
\author{Alex J. Matthies}
\affiliation{Department of Physics, Durham University, South Road, Durham DH1 3LE, United Kingdom}
\author{Luke Caldwell}
\affiliation{Department of Physics and Astronomy, University College London, Gower Street, WC1E 6BT, London, UK}
\author{Hannah J. Williams}
\email{hannah.williams4@durham.ac.uk}
\affiliation{Department of Physics, Durham University, South Road, Durham DH1 3LE, United Kingdom}

\begin{abstract}
Optical tweezers are a powerful tool for creating defect-free arrays of atoms and molecules, enabling advances in quantum simulation, computation, and precision metrology. However, the achievable array size is limited by the initial loading fraction, typically $50\,\%$ for atoms and $35\,\%$ for molecules. Here, we propose a general scheme for enabling multiple loading cycles by protecting trapped particles using a repulsive barrier. We show that collision-limited lifetimes of particles in protected tweezers can exceed one second, leading to filling fractions of over $80\%$ after four loading cycles. Combined with existing rearrangement techniques, this approach enables efficient unity filling of tweezer arrays and provides a scalable pathway towards larger quantum technology platforms.

\end{abstract}

\maketitle

\section{Introduction}
\begin{figure*}
\centering
\includegraphics[width=\linewidth]{rearrangment11.png}
\caption{\label{fig:Rearrangement} (a) Illustration of the barricade tweezer set up. Spatial light modulators (SLMs) are used to control the phase patterns for both the attractive and repulsive beams, before they are combined and focused through the tweezer objective. The barricade potential experienced by a protected particle is shown in the inset. (b) An experimental sequence for two full loading cycles, with 1D potential plots demonstrating the protection mechanism for an example array of three tweezers at different points in the cycle. The colored blocks indicate the timings for the cooling (teal), imaging (pink) and repulsive (blue) light. }
\end{figure*}

Arrays of optical tweezers provide a powerful platform for quantum science~\cite{Browaeys2020,Kaufman2021}, enabling the trapping and control of individual atoms~\cite{Endres2016,Barredo2016,Lee2016,Norcia2018,Saskin2019,Jackson2020, Sheng2022} or molecules~\cite{Anderegg2019Science,bao2023spinExchange,Holland2023a,Park2023,Vilas2024,Ruttley2025LongLivedEntanglement}. These arrays can be generated in arbitrary geometries~\cite{Nogrette2014,Kim2016,Schymik2020}, dynamically reconfigured~\cite{Bluvstein2022,Holland2023entanglementArray}, and scaled to thousands of sites~\cite{Manetsch2025Tweezer6100,Pichard2024,Chiu2025}. In combination with tunable interactions between atoms or molecules, these capabilities have enabled pioneering advances in quantum simulation~\cite{ebadi2021quantum,Scholl2021Quantum2DAFM}, quantum  computation~\cite{Graham2022NeutralAtomAlgorithms,Evered2023ParallelGates,Bluvstein2025} and metrology~\cite{Madjarov2019,Young2020}. All these applications rely on defect-free arrays, which, as we discuss below, become increasingly challenging to achieve with increasing array size.

Optical tweezers are generally loaded from dilute gases, where laser cooling dissipates kinetic energy and enables atoms or molecules to be loaded into conservative traps. The presence of this cooling light causes light-assisted collisions, limiting the maximum occupation of a tweezer to one particle~\cite{Schlosser2002}. This stochastic process leads to the single-site loading probability saturating at $50\,\%$ for atoms~\cite{Schlosser2001SubPoisson,Schlosser2002}. For molecules, the experimentally observed loading fractions are lower, reaching only approximately $35\,\%$ ~\cite{Anderegg2019Science,Holland2023entanglementArray}. A method to exceed the stochastic limit is to engineer the collisions such that only one particle is ejected~\cite{Grunzweig2010,Lester2015}. In this way, array-averaged filling fractions of up to $80\,\%$ have been achieved for alkali atoms ~\cite{Aliyu2021,Brown2019,Angonga2022} and up to $93\,\%$ for Yb~\cite{Jenkins2022}. Extending this approach to molecules, however, appears to be infeasible due to incompatibility with collisional shielding~\cite{Walraven2024PRL}.

A widely adopted strategy to achieve defect-free arrays is direct rearrangement of the particles. Here, loaded particles are shuttled into a target configuration by moving tweezers~\cite{Barredo2016,Endres2016,Lee2016,Schreck2025SLMrearangment}. Such rearrangement procedures scale poorly with increasing array size, and ultimately the size of the fully-filled array is set by the initial filling fraction. Together, these considerations strongly motivate the development of techniques to enhance loading fidelities of tweezer arrays.

A promising approach is to utilize multiple loading cycles, which cumulatively increase the overall filling fraction. To achieve this, already loaded particles must remain trapped in the presence of the cooling light. Recently, two schemes have been proposed which involve transferring trapped particles into dark states, thereby protecting them during further loading cycles~\cite{Shaw2023,Walraven2024PRL}. Both approaches require the availability of long-lived shelving states and high-fidelity state preparation and transfer. In this work, we introduce a general protection protocol whereby loaded sites are identified and the trapped particles are protected by the addition of a repulsive barricade which prevents further particles from entering the trap.

The barricade potential is realized by superimposing two optical tweezers at different wavelengths. Figure ~\ref{fig:Rearrangement}(a) shows the overlapping of the trapping tweezer, red detuned from an optical transition (red path), with a blue-detuned beam (blue path) which generates a repulsive potential. By carefully selecting the relative beam waists and powers, a potential can be engineered to have a repulsive barrier surrounding a trapping core. High-speed spatial light modulators (SLMs)~\cite{Schreck2025SLMrearangment,Vanackere2025} are used to control the parameters of the two focused beams.  
The protection sequence is sketched in Fig.~\ref{fig:Rearrangement}(b). Particles are initially cooled into tweezers, then the array is imaged and loaded sites are identified. 
The relevant repulsive potentials are then ramped on adiabatically, to avoid parametric heating effects, and the power of the trapping tweezers are simultaneously increased to maintain a constant trap depth. Unoccupied tweezers remain unmodified, to be loaded in subsequent cycles. 

The improvement in loading efficiency directly follows from the cumulative loading over $N$ cycles, given by,
\begin{equation}
    \label{eqn:fillfrac} 
    \eta = \eta_0\cdot\frac{1-(e^{-t/\tau}+\eta(\tau)-\eta_0)^N}{1-e^{-t/\tau}-\eta(\tau)+\eta_0 },
\end{equation}
 where $\eta_{\rm0}$ is the initial filling fraction,  $t$ is the duration of each loading cycle,  $\tau$ is the collisional lifetime of a protected particle and $\eta(\tau) = \eta_0(1-e^{-t/\tau})^2$ is the probability of reloading a particle that has been lost from a protected site in the same cycle (see Appendix). For atomic and molecular tweezers with $\eta_0 = 50\%$ and $35\%$ respectively, in the limit of no collisional loss ($\tau \rightarrow \infty$), three additional loading cycles (N = 4) gives theoretical maximum total loading efficiencies of $\eta = 94\%$ and $82\%$, which are comparable to the dark-state protection schemes~\cite{Shaw2023,Walraven2024PRL}.
 
\section{Methods\label{Methods}}

The barricade potential, $U_{\rm b}(\vec{r}) = U_{\rm att}(\vec{r})+U_{\rm rep}(\vec{r})$, is the sum of the light shifts from the attractive and repulsive beams. The scalar light shift for a particle in an off-resonant optical beam is,
\begin{equation}
    U(\vec{r})= -\frac{\alpha_{\lambda}}{2 \epsilon_{\rm 0} c} I(\vec{r}),
\end{equation}
where $I(\vec{r})$ is the intensity of the applied field and $\alpha_{\rm\lambda}$ is the real part of the polarizability\footnote{Tensor light shifts may also be important depending on the species being trapped.}. The sign of the polarizability sets the sign of the force, for a simplified two-level system, light that is red-detuned from the transition results in an attractive potential ($\alpha_{\lambda}>0, U<0$), while light that is blue-detuned creates a repulsive potential ($\alpha_{\lambda}<0$, $U > 0$). 

Optical tweezers are created by focusing light through a lens with a high numerical aperture (NA). We calculate the intensity distribution by modeling the electric field near the focus via a non-paraxial vectorial model~\cite{book_opticsF2F}. The vector field at a point $\vec{r}= (x,y,z)$ near the focus is
\begin{equation}
\label{eqn:Richards-Wolf}
    \vec{E}(\vec{r}) = -\frac{ik}{2\pi}\iint_{\rm \Omega}\vec{a} e^{ik(\vec{s}\cdot\vec{r})}\rm d\Omega,
\end{equation}
where $k$ is the wave vector, $\vec{s}$ is a vector on the unit sphere, and $\vec{a}$ is the amplitude function of the light before the objective lens, containing information on the input polarization and NA of the lens. The integration is over $\Omega$, the solid angle subtended by the lens at the focus. Numerical evaluation of this expression yields the focal intensity $I(\vec{r}) = \frac{1}{2} \epsilon_0 c |\vec{E}(\vec{r})|^2$. The full barricade potential $U_{\rm b}(\vec{r})$ is then evaluated, from which information on the trap depth, barrier height and shell radius can be extracted.

\begin{figure}
    \centering
    \includegraphics[width=1\linewidth]{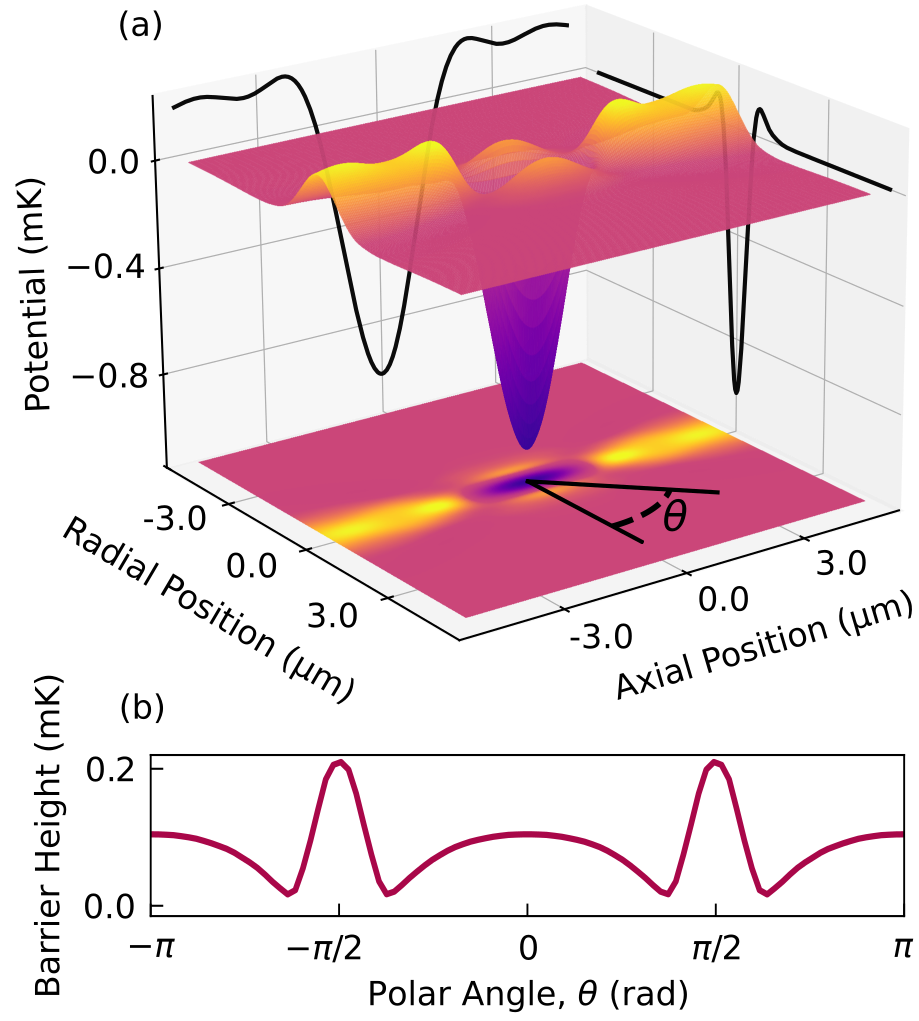}
    \caption{(a) Optical potential, $U_{\rm b}$, formed by superimposing an attractive tweezer ($\lambda_{\rm att} = 780$\,nm, $w_{att} = 0.55~\upmu$m and $P_{\rm att} = 7.5$\,mW) with a repulsive beam ($\lambda_{\rm rep} = 508$\,nm, $w_{rep} = 1.1~\upmu$m and $P_{\rm rep} = 5.5$\,mW) for the $\ket{F, m_{\rm F}} = \ket{2,\pm2}$ level of the $X^2\Sigma (N=1)$ state of CaF. 
    Dark purple regions indicate the trapping core, while orange regions indicates the repulsive barrier. Black lines show 1D axial and radial slices through the trap center. (b)  Plot showing the polar angular dependence, $\theta$, of the maximum barrier height $U_b^{\mathrm{max}}(\theta, \phi =0)$.}
    \label{fig:Potential}
\end{figure}

The combined potential landscape of the two beams depends on the wavelength-dependent polarizability of the trapped particle, and the beam waists and powers. The wavelengths are chosen with consideration of both the scattering rate and polarizability.
Provided the waist of the repulsive beam $w_{\rm rep}$ is larger than that of the attractive beam $w_{\rm  att}$, a barrier can be created, where $w_{\rm att}$ is typically set to be diffraction limited. A range of powers and $w_{\rm rep}$ can then be explored and values selected based on available powers and desired barrier height and trap depth. Figure~\ref{fig:Potential}(a) shows an illustrative plot of a barricade potential showing a clear central trapping region surrounded by an anisotropic barrier. In Fig.~\ref{fig:Potential}(b) we plot the maximum barrier height $U_b^{\mathrm{max}}(\theta,\phi= 0) \equiv U_b^{\mathrm{max}}(\theta)$ as a function of polar angle $\theta$, which shows the strong anisotropy. The minimum barrier height will ultimately limit the protection of the core, however, provided there is a barrier in all directions, repulsion will still occur. 

We investigate the effectiveness of the barricade at preventing particles from entering the trapping core for different barrier heights. Due to the anisotropy of the barricade potential, we model the loading dynamics using 3D semi-classical Monte-Carlo trajectory simulations. An ensemble of particles with mass $m$ are initialized with velocities drawn from a Maxwell-Boltzmann distribution at temperature $T_0$, with number density $n_{\rm 0}$. We vary the barrier height by changing the powers of the two beams for a chosen $w_{\rm rep}$, and calculate the full three-dimensional potential experienced by the particle consisting of a single tweezer summed with an external reservoir potential\footnote{The reservoir is a large harmonic potential, with trap frequency three orders of magnitude smaller than the tweezer, which maintains particle density during the simulations.}. 
The trajectories of the particles are evolved according to $m\ddot{\vec{r}} = - \nabla U_{\rm tot}(\vec{r})$. From the simulations we extract a flux for particles entering the trapping core, defined by $R_{\rm core}$. We present two example cases in the next section which demonstrate that a suppression in flux by several orders of magnitude can be achieved in the presence of the barricade potential.

\section{Results}
\begin{figure*}
    \centering
    \includegraphics[width=2\columnwidth]{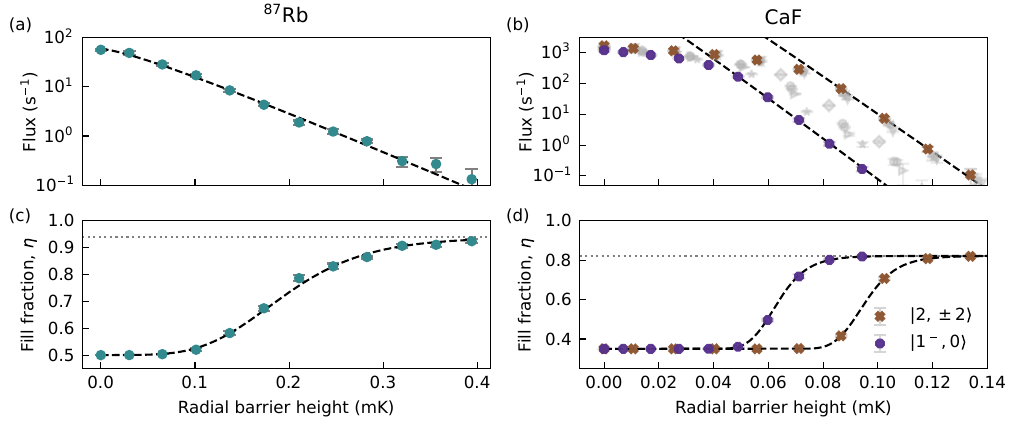}
    \caption{(a,b) Simulated particle flux through the core of the barricade tweezer as a function of radial barrier height for (a) $^{87}$Rb and (b) CaF. (c,d) Cumulative fill fraction of the array, $\eta$, as a function of radial barrier height after $N=4$ loading cycles, calculated using Eq.~\ref{eqn:fillfrac} for (c) $^{87}$Rb and (d) CaF. The black dashed lines show the fit of the isotropic barrier model, see Appendix. The grey dotted lines shows the maximum fill fraction for $N=4$ in the limit of $t\ll\tau$. In the CaF case, the $\ket{F, m_{\rm F}} = \ket{1^-,0}$\,$^2$\,(purple octagons) and $\ket{2,\pm2}$ (brown crosses) correspond to the ``best" and ``worst" case barriers.}
    \label{fig:Loading}
\end{figure*}

We evaluate the barricade protocol for two representative species, routinely trapped in optical tweezers,  $^{87}$Rb and CaF. We follow the procedure outlined above to calculate the flux through the trapping core for different barrier heights. The results are presented versus the radial barrier height, at $\theta=\phi = 0$, as a representative value for the anisotropic potential experienced by the particle.

\subsection{Rb}
Rubidium atoms are typically loaded into optical tweezers directly from a MOT. We use simulation parameters of temperature $T_{\rm 0}=59\,\upmu\mathrm{K}$ and peak density $n_{\rm 0}=5\times10^8$\,cm$^{-3}$, consistent with experimental values. We calculate the attractive tweezer potential for $\lambda_{\rm att} = 820$~nm light focused to a $w_{\rm att}=0.79~\upmu$m, and set $U_{\rm att}(0)=1$~mK~\cite{Nogrette2014}. For the repulsive beam, we select $\lambda_{\rm rep} = 759$~nm and $w_{\rm rep}=1.2~\upmu$m.

Using the methods described, we explore barrier heights up to $U_{\rm b}^{\mathrm{max}}\left(\theta = 0\right)=k_{\rm B}\times390~\upmu$K. To calculate the flux, we consider that an atom has entered the trapping region, and hence would either be trapped or cause a light-assisted collision, if it gets to within $ R_{\rm core} = 50$\,nm, chosen to be representative of the Condon radius for rubidium~\cite{Fung2016Technologies,Fuhrmanek2012}. Figure~\ref{fig:Loading}(a) shows the simulated suppression of incident flux versus radial barrier height. The decrease in flux scales almost exponentially with increasing radial barrier height. This behavior is expected for particles drawn from a thermal reservoir experiencing an isotropic barrier. Over the range of parameters explored here, the barricade potential exhibits only weak angular variation around the trapping core. Consequently, the barricade potential can be approximated by an isotropic barrier characterized by the radial barrier height, and the resulting flux suppression is well described by an analytical isotropic barrier model (see Appendix), shown as the black dashed line in Fig. 3(a).

A barrier height of $U_{\rm b}^{\mathrm{max}}\left( 0\right)=k_{\rm B}\times280~\upmu$K, suppresses the flux to $0.78(6)~$s$^{-1}$ corresponding to $P_{\rm rep} = 2.7$\,mW and $P_{\rm att} = 2.7$\,mW. These powers are comparable with those typically used for optical tweezers, making this technique compatible with large arrays. With this flux, the collisional lifetime of an atom in the protected tweezer would be $\tau =1.3(1)$\,s. 

To calculate the cumulative filling fraction for multiple loading cycles we need to assign a cycle time $t$, during which cooling light is present and thus expulsion from a trap can occur. We choose a loading time of 70~ms, such that the Poisson statistics have reached the saturation limit of $\eta_0=0.5$, and an imaging time of 5~ms, giving a total cycle time of $t=75$~ms. Another potential source of loss is due to off-resonant scattering, and hence heating from the blue-detuned light~\cite{Grimm2000DipoleTraps}. We estimate that the presence of this light leads to an off-resonant scattering rate of $9~\text{s}^{-1}$ and a heating rate of $\dot{T} = 3~\upmu\text{K/s}$, which causes negligible heating when compared to the 1\,mK trap depth. Using Eq.~\ref{eqn:fillfrac}, we find a cumulative filling fraction of $\eta = 87\%$ for $N=4$. To achieve higher fill fractions, higher barriers or more cycles would be needed. Fig.~\ref{fig:Loading}(c) shows the fill fraction $\eta$ as a function of the barrier height for $N=4$ loading cycles. As the radial barrier height is increased $\eta$ rises steeply before saturating at the theoretical maximum of 94\%. Importantly, this saturation occurs at barrier heights readily achievable with standard tweezer powers, demonstrating that the theoretical maximum is practically accessible without significant additional optical power.

\subsection{CaF}

The state-of-the-art filling fraction for molecular tweezer arrays remains substantially below the atomic stochastic limit, saturating at around 35\%. Protocols that enable multiple loading cycles therefore offer a particularly large improvement for molecules, making CaF a highly relevant test of the barricade protocol. 

Molecular MOTs reach number densities several orders of magnitude lower than alkali MOTs~\cite{Norrgard2016,Williams_2017,Anderegg2017,Collopy2018}. As a result, CaF tweezers are not loaded directly from a MOT but instead from an ODT, where much higher densities and lower temperatures are obtained. 
In our simulations we use routinely achievable loading parameters of $T_0=9~\upmu$K and $n_0=1\times10^9~\text{cm}^{-3}$, with molecules held in a $100~\upmu$K deep ODT.
For the attractive tweezer we use $\lambda_{\rm att}=780$\,nm, focused to a waist of $w_{\rm att} = 0.55~\upmu$m to produce a trap of depth 
$U_{\rm att}(0)=1$~mK, which is close to the properties used in experiments~\cite{Anderegg2019Science,Holland2023a}.
For the repulsive potential we select $\lambda_{\rm rep} = 508$\,nm, and $w_{\rm  rep}=1.1\,\upmu$m. It is important to note that the tensor polarizability must be considered for molecules~\cite{Humphreys2025}, and as such the barricade potential now depends on the $\ket{F, m_{\rm F}}$ state. To capture this, we run simulations for each $m_F$ potential individually. In practice, MHz-rate photon scattering during imaging and cooling causes a molecule to traverse the full $m_{\rm F}$ manifold many times while crossing the barrier, and it is therefore conservative to treat the worst-case $m_F$ state as a lower bound on the effectiveness of the barricade.

Since an effective Condon radius for CaF is unknown, we use a conservative definition for the core, $R_{\rm core} = r_{\rm FWHM} = 0.63~\upmu\text{m}$, where $r_{\rm FWHM}$ is the full width at half maximum of the attractive tweezer intensity profile. The resulting flux versus radial barrier height is shown in Fig.~\ref{fig:Loading}(b). The gray points show the flux for different $m_{\rm F}$ states, with the $|F, m_{\rm F}\rangle = |2, \pm2\rangle$ and $|1^{-}, 0\rangle$~\footnote{\label{fn:caf}The superscript in the $ F = 1^{-}$ state indicates that it is the lower energy of the two $F = 1$ states in ground-state manifold of CaF.} states highlighted in brown and purple, respectively, as worst and best case scenarios in terms of power requirements. Two distinct scaling regimes are visible. At low repulsive power, strong angular structure in the barricade potential forms narrow funnels that guide particles into the core for specific incident angles. As the repulsive power is increased, these funnels close and the barricade becomes fully repulsive. The radial barrier height then increases monotonically with optical power, and the flux suppression is well described by the isotropic barrier model (see Appendix), shown by the black dashed lines in Fig.~\ref{fig:Loading}(b) for the two extreme cases. The transition point at which the Monte Carlo results converge to the isotropic model is itself $m_{\rm F}$-dependent,  with the $\ket{1^-,0}$ state converging at a lower barrier height than 
$\ket{2,\pm2}$. 
 

 To look at multiple loading cycles, we consider a cycle time of 15\,ms, comprising of 10\,ms of cooling and 5\,ms of imaging, during which light-assisted collisions occur. The blue-detuned light leads to a maximum scattering-induced heating rate of $\dot{T} = 4.3~\upmu \text{K}/\text{s}$~\cite{Grimm2000DipoleTraps}, which causes negligible heating in comparison to the 1\,mK trap depth. Figure~\ref{fig:Loading}(d) shows $\eta$ versus the radial barrier height for both the $|2, \pm2\rangle$ and $|1^{-}, 0\rangle$ states for $N=4$. At low barrier heights, the effect of the attractive funnels dominates and no increase in $\eta$ is observed. At large barrier heights, both curves saturate at the theoretical maximum of $82\,\%$. A barrier height of $U_{\rm b}^{\mathrm{max}}\left( 0\right)=k_{\rm B}\times120~\mu$K, corresponding to $P_{\rm att}=7.7$\,mW and $P_{\rm rep}=6.3$\,mW, suppresses the flux by more than three orders of magnitude, to 0.74(15)\,$\text{s}^{-1}$ for molecules in $|2, \pm2\rangle$ and a collision-limited lifetime of $1.3(3)$\,s leading to a cumulative fill fraction of $\eta=81\%$ after four loading cycles. Since the molecules will explore the full ground-state manifold, and the true Condon radius for CaF is unknown and likely smaller than the $R_{\rm core} = r_{\rm FWHM}$ assumed here, these power requirements are conservative and in practice lower powers should achieve the same collision-limited lifetime.

\section{Conclusion}
We have introduced the barricade tweezer that enables a near-deterministic method of loading arrays of optical tweezers through repeated loading cycles. By superimposing attractive and repulsive optical potentials, a barricade is formed which protects loaded traps while unoccupied sites remain available for loading. The resulting enhancement of the fill fraction is independent of array size, and the scheme is applicable to any species for which attractive and repulsive potentials can be created.

From our simulations we identified realistic trap parameters which would achieve collision-limited lifetimes of $1.3(3)$\,s for both $^{87}$Rb and CaF. Applying four loading cycles with these lifetimes, we predict cumulative filling fractions of 87\% for $^{87}$Rb and 81\% for CaF. Crucially, the filling fraction saturates rapidly with barrier height, making the theoretical maxima for four cycles practically accessible. Combined with parallel rearrangement techniques, this protocol provides a scalable route towards preparation of large, defect-free arrays of atoms and molecules for quantum computation and simulation.

\section{Acknowledgments}
We would like to thank Antoine Browaeys and Pasqal for initial discussion around the concept of the barricade tweezers, Ifan Hughes for discussion on calculating optical potentials, Sam Walker for aiding with the simulation of bulk gases, Bethan Humphreys, Becca J. Hedley, and members of the Centre for Cold Matter at Imperial College London for providing feedback on the manuscript.  HJW acknowledges UKRI grant MR/X033430/1 and LC acknowledges UKRI grant MR/Y017056/1. This work has made use of the Hamilton HPC Service of Durham University.

\section{Data Availability Statement}
Data and code are available on GitHub in the repo: \href{https://github.com/ArchieB151/barricade-tweezers}{ArchieB151/barricade-tweezers}.

\bibliography{ref2}
\appendix

\section{Appendix: Fill Fraction Model}
We derive the cumulative fill fraction $\eta$ after $N$ loading cycles by considering the probability $P_{N}$ that a given tweezer site is occupied at the end of the $N$-th loading cycle. In the first cycle, no barricade is present and thus $P_{\rm 1} = \eta_{\rm 0}$. In subsequent cycles, three processes contribute to the fill fraction. A protected particle survives a subsequent loading cycle of duration $t$ with probability $e^{-t/\tau}$; an unoccupied site is loaded with probability $\eta_{\rm 0}$; and finally a tweezer site that underwent loss is reloaded within the same cycle with probability $\eta(\tau)$. This gives the recursion relation,
\begin{equation}
\begin{split}
    P_{N} = \; & P_{N-1}\,e^{-t/\tau} + \left(1 - P_{N-1}\right)\eta_{\rm 0} + P_{N-1}\eta(\tau).
\end{split}
    \label{eqn:recursion}
\end{equation}

Solving this equation as a linear recurrence with constant coefficients, with the initial condition  yields Eq.~\eqref{eqn:fillfrac} of the main text. 

For atoms, the pairwise ejection effect resulting from light-assisted collisions is the dominant mechanism during loading. We model the number of particles entering the core as a Poisson process with mean $\lambda = t/\tau$. 
A particle in a protected site is lost and subsequently reloaded within the same cycle if and only if an even, non-zero number of particles enter the core. This joint probability is
\begin{equation}
\label{eqn:eta_tau}
    \eta(\tau) = \frac{1}{2}\!\left(1 - e^{-t/\tau}\right)^2,
\end{equation}
which recovers the saturation limit of $\eta_0 = 0.5$ in the limit of $t/\tau\gg 1$. For molecules, the same pairwise ejection mechanism is expected to operate, yet the experimentally observed fill fraction saturates at approximately $35\,\%$,  suggesting  an additional loss mechanism, the nature of which remains unknown. In the absence of a model for this additional loss, we adopt a heuristic approach and generalize Eq.~\ref{eqn:eta_tau} to
\begin{equation}
    \eta(\tau) = \eta_{\rm 0}\!\left(1 - e^{-t/\tau}\right)^2,
\end{equation}
which preserves the pairwise ejection effect, captures the observed saturation for both species, and reduces to Eq.~\ref{eqn:eta_tau} when $\eta_0 = 0.5$.

\section{Appendix: Isotropic Barrier Model}

We derive an analytical expression for the flux suppression factor $S$ in the limit of an isotropic barrier, of height $U_{\rm b}$. In this limit, the barrier is spherically symmetric, so a particle 
can enter the trapping core only if its total kinetic energy $E$ exceeds 
$U_{\rm b}$. The suppression factor $S$ is then the fraction of the 
thermal ensemble that satisfies this condition.

For a gas in thermal equilibrium at temperature $T$, the 
Maxwell-Boltzmann energy distribution is
\begin{equation}
    g(E) = 2\sqrt{\frac{E}{\pi}}\left(\frac{1}{k_{\rm B} T}\right)^{3/2} 
    \exp\!\left(-\frac{E}{k_{\rm B} T}\right).
\end{equation}
Integrating over all particles energetic enough to surmount 
the barrier gives
\begin{align}
\label{Eqn:SuppresionFactor}
    S(T,U_{\rm b}) &= \int_{U_{\rm b}}^{\infty} g(E)\, dE \nonumber \\
      &= \operatorname{erfc}\!\left(\sqrt{\frac{U_{\rm b}}{T}}\right) 
         + \frac{2}{\sqrt{\pi}}\sqrt{\frac{U_{\rm b}}{T}}
           \exp\!\left(-\frac{U_{\rm b}}{T}\right).
\end{align}

The isotropic model used in Fig.~\ref{fig:Loading}(a) and (b), takes the form $F(T,U_{\rm b},j_0)=j_0S(T,U_{\rm b})$ where $j_0$ is the flux in the absence of a barrier and both $j_0$ and $T$ are free parameters in the fit. The model is valid when the angular variation in the barrier height is small compared to $k_{\rm B} T$. For $^{87}$Rb this condition holds across the full range of barrier heights explored, and the Monte Carlo results are well described by Eq.~\ref{Eqn:SuppresionFactor}, as seen in Fig.~\ref{fig:Loading}(a). For CaF, the selected wavelengths and beam waists introduce strong angular variation in the barrier height. At low barrier heights this gives rise to narrow funnels through which particles can reach the trapping core at specific incident angles. This behavior is not captured by the isotropic model, resulting in a deviation from 
Eq.~\ref{Eqn:SuppresionFactor} at low barrier heights. Once the repulsive power is sufficiently large that protection exists at all angles, the funnels vanish and the Monte Carlo results are well described by Eq.~\ref{Eqn:SuppresionFactor}. As such, the model is only fitted to the high barrier regime resulting in the fits shown in Fig.~\ref{fig:Loading}(b). Additionally, the form of the barriers depends on the $m_F$ state, this presents as an effective temperature dependence when fitting Eq.~\ref{Eqn:SuppresionFactor} to the data.

\end{document}